# The program Simourg for simulating the response functions of gamma detectors with simple geometries

V. V. Kobychev[1,2]

[1] *Institute for Nuclear Research, NAS of Ukraine, Kyiv, Ukraine*
[2] *Department of Physics and Astronomy, Seoul National University, Seoul, Korea*

The program Simourg (Simulator of Usually Requested Geometries) is based on the Geant4 toolkit and created for Monte Carlo simulation of gamma-ray spectrometric nuclear detectors with a simple axial symmetric geometry, which is typical for many tasks of studying the decay of long-lived nuclei and measuring the radioactivity of natural objects. The program is designed for quick estimation of the effectiveness and the response function of the detector to monoenergetic gamma quanta in the energy range from keV to several MeV.

1. Introduction.

Measurements of the absolute activity of gamma sources require knowledge of the efficiency and response function of the detector used. These parameters can be measured experimentally using calibration sources at several discrete gamma-ray energies. However, at the stage of experiment planning and design of the measuring setup, experimental measurements of the response function are not yet available. When the geometry of the setup is changed, efficiency calibration must be repeated. The placement of a calibration source in the setup usually differs (to a greater or lesser extent) from the placement of the measured sample, meaning the efficiencies may also differ. The problems listed above are solved by computer Monte Carlo simulation of the experimental setup. There exist a number of software packages and libraries for numerical simulation of particle transport and Monte Carlo modeling of the response functions of nuclear-physics detectors. Examples include Geant4, EGS4, MCNP, etc. However, working with them requires the user — in addition to understanding the physical processes occurring in the detector — considerable effort to study the specific features of the package and the programming language in which it is implemented. Modeling a new detector or even making small changes to the geometric and physical parameters of an existing model generally requires recompiling the pro-gram. In practice, however, for detectors with simple geometry (i.e. in the majority of cases), the user does not need many of the advanced capabilities built into the above-mentioned packages.

In experimental non-accelerator particle physics (as well as in studies of rare nuclear processes and measurements of low-level activities, including gamma activity of natural objects), a typical task is to construct the response function of a spectrometric detector (e.g., scintillation, semiconductor, or cryogenic bolometric) to gamma rays with energies up to several MeV emitted by a source located inside shielding near the detector. Another typical task in these applications is to determine the detector response to background gamma rays emitted by a source uniformly distributed in one of the structural elements of the setup (for example, in photomultiplier glass or in the shielding). In such setups the elements usually possess axially symmetric geometry and are coaxial.

For these reasons, the task was set to create a Monte Carlo program that would allow the user, with minimal effort, to build a model of a simple gamma detector and calculate its response function to a monoenergetic gamma source. An example of such a simple detector is a cylindrical scintillator located inside massive shielding and irradiated by gamma rays from a point-like or cylindrical source. User interaction with the program is reduced to preparing a simple text configuration file containing the geometric characteristics of the detector components (dimensions, coordinates), their physical properties (chemical composition, density), the location of the gamma source (point-like or extended), the gamma-ray energy, and a number of auxiliary parameters related to the output of results.

2. Characteristics of the developed software.

To solve this task, the program Simourg (*Simulator of Usually Requested Geometries*) was developed. The program is based on the Geant4 class library [1] developed at CERN for Monte Carlo simulation of nuclear-physics experiments; Simourg version v1.0 uses Geant4 version 9.2.patch-01. The program can run on all platforms supported by Geant4 (Linux, MacOSX, SunOS, and MS Windows XP/Vista). For the first three platforms, Geant4 must be pre-installed and the Simourg source code compiled with the Geant4 libraries; for Windows, a pre-compiled executable is available for download. The program, its source code, and documentation can be downloaded from http://lpd.kinr.kiev.ua/kobychev/Simourg. During execution, the program uses the G4EMLOW library (version 6.3 or later required) for low-energy electromagnetic interaction data. This library is freely distributed by CERN [2] and must be installed on the user's computer; Simourg looks for it using the path specified in the environment variable G4LEDATA. When running under Windows, Microsoft .NET Framework version 3.0 or higher is required (freely downloadable from the manufacturer's website [3]), although in most cases it is already installed because many applications use it.

The model includes only electromagnetic interactions of photons, electrons, and positrons in matter: Compton and Rayleigh scattering, bremsstrahlung, electron–positron pair production, positron annihilation, photoelectric effect, and ionization (followed by emission of Auger electrons and fluorescence photons). Cross sections of inelastic nuclear reactions are negligible at the energies considered and are not taken into account.

3. Features of operation.

To run the program, the user should define in a text configuration file at least one physical volume (the detector itself), the location of the gamma source, and the gamma energy. In addition, some auxiliary parameters related to result output are specified. An example of a "minimal" configuration file is shown in Fig. 1.

```
/user/reset                                 # Parameters of the simulation
# Parameters of the Detector volume
/user/D_Det   50.00 mm                      /user/EGamma 2614.5 keV
/user/ZL_Det 40.00 mm                       /user/FWHM1 0.0
/user/Z_Det   0.00 mm                       /user/FWHM2 2.0
/user/MaterialDetDensity 8.0 g/cm3          /user/Threshold 1.0 keV
/user/MaterialDetElementName Cd             /user/ELowLimit 1.0 keV
/user/MaterialDetFormulaNum    1            /user/DELowLimit 1000.0 nm
/user/MaterialDetElementName W              /user/DGLowLimit 1000.0 nm
/user/MaterialDetFormulaNum    1            /user/ChannelWidth 1.0 keV
/user/MaterialDetElementName O              /user/numberOfRuns 1000000
/user/MaterialDetFormulaNum    4            /user/Step 1000
                                            /user/RandomSeed 90115037
# Parameters of the Source volume           /user/WaitCommand 0
/user/XL_Src 1.00 mm                        /user/VerboseAll 0
/user/YL_Src 1.00 mm                        /user/DoPicture 0
/user/ZL_Src 1.00 mm
/user/Z_Src 3.50 cm                         /user/showAll
```

Fig. 1. Example of a minimal configuration file.

In the configuration file shown, the volume `Det` is a cylindrical CdWO4 detector with diameter `D_Det = 50 mm` and height `ZL_Det = 40 mm`, located at the origin (`Z_Det = 0 mm`). The material density is `MaterialDetDensity = 8.0 g/cm³`. The chemical composition is specified by the chemical formula with the number of atoms of each element (up to 20 elements allowed). Since surrounding volumes are not explicitly defined, they are considered vacuum by default. The source is a virtual volume `Src` (i.e., it has no physical material) shaped as a cube with side lengths `XL_Src = YL_Src = ZL_Src = 1 mm`, within which the initial vertices of gamma rays are uniformly distributed with isotropically random initial directions. The source center is located on the Z-axis at `Z_Src = 3.5 cm` from the detector center (coordinates `X_Src` and `Y_Src` are not specified and default to zero; however, unlike other volumes, the source center is not required to lie on the Z-axis).

In this example, the program will generate `numberOfRuns = 1 000 000` gamma rays with energy `EGamma = 2614.5 keV` and isotropically distributed initial directions. The spectrum of energy deposited in the detector is written to the file `SpectrumRaw.dat`; the channel width is set by the parameter `ChannelWidth` (1 keV in the example). This file can be used, among other things, to determine the full-energy peak efficiency. If the energy dependence of the detector resolution is known (usually of the form

$$R(E) = (a_1 + a_2 \cdot E)^{1/2}, \qquad (1)$$

where $E$ is energy, $R$ is FWHM resolution), the user can specify parameters $a_1$ and $a_2$ (as `FWHM1`, keV², and `FWHM2`, keV), and the program will produce a resolution-broadened spectrum in the file `SpectrumBlur.dat`. Before writing to this file, the deposited energy of each event with non-zero energy deposition is shifted by a random Gaussian-distributed value with zero mean and energy-dependent width according to Eq. (1). The spectrum files are two-column ASCII files; each line corresponds to one energy channel (containing the channel centre energy and the number of counts).

Each line in the configuration file is a command that sets a particular parameter or performs an action. A complete list of user commands and their description is provided in the file `help.txt` supplied with the program.

In addition to the mandatory volumes (`Det` and virtual `Src`), the user may optionally define several additional physical volumes by specifying their diameters, central coordinates, density, and material composition.

- `Core` – a shell surrounding the `Det` volume. If it is defined, `Det` must be completely immersed in it, otherwise an error is reported. If `Core` parameters are not specified, it is automatically created with dimensions 0.001 mm larger than `Det`, centered at the same point, and filled with the same material as the shield (see below). Thus, when not explicitly defined, it has practically no effect. This volume can describe, for example, a dead layer of a semiconductor detector or a cavity in the shielding.
- `Shield` – passive shielding, the outermost volume; all other volumes must lie inside it, otherwise an error is issued. If not defined, it is created large enough to contain all other volumes, centered at the origin, and filled with vacuum by default. Particles escaping this volume are killed.

- `Cavity` – a volume inside `Det`. If it protrudes from `Det`, an error is reported. Intended for internal wells typical of HPGe detectors or scintillators with wells. It is not created if not described.
- `Top1, Top2, Top3, Bot1, Bot2, Bot3` – intended for photomultipliers, light guides, cold fingers, etc. They are located outside `Core` but inside `Shield`. Overlap with any other physical volume (except `Shield`) causes an error. It is created only if explicitly defined.

The hierarchy of physical volume nesting is shown as a directed tree graph in Fig. 2; the arrows indicate containment. All volumes except the virtual `Src` must be coaxial with the common Z-axis. All volumes are cylinders, except `Src`, `Core`, and `Det`, which may also be rectangular parallelepipeds (in the latter case, instead of diameter, e.g., `D_Src`, the user specifies `XL_Src` and `YL_Src`). Overlap of any two physical volumes is forbidden. The virtual `Src` may be placed arbitrarily and may coincide with or intersect physical volumes (since it has no material). However, it must lie completely inside `Shield`. Fig. 3 shows an example geometry that includes all allowed physical volumes.

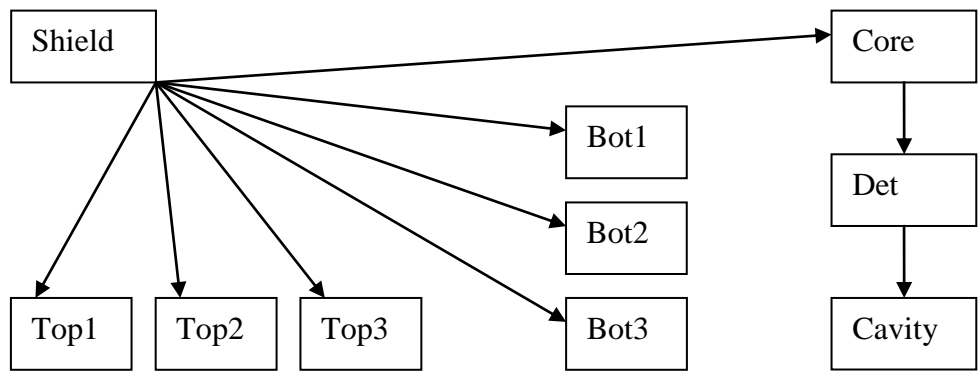

Fig. 2. Hierarchy of physical volume nesting.

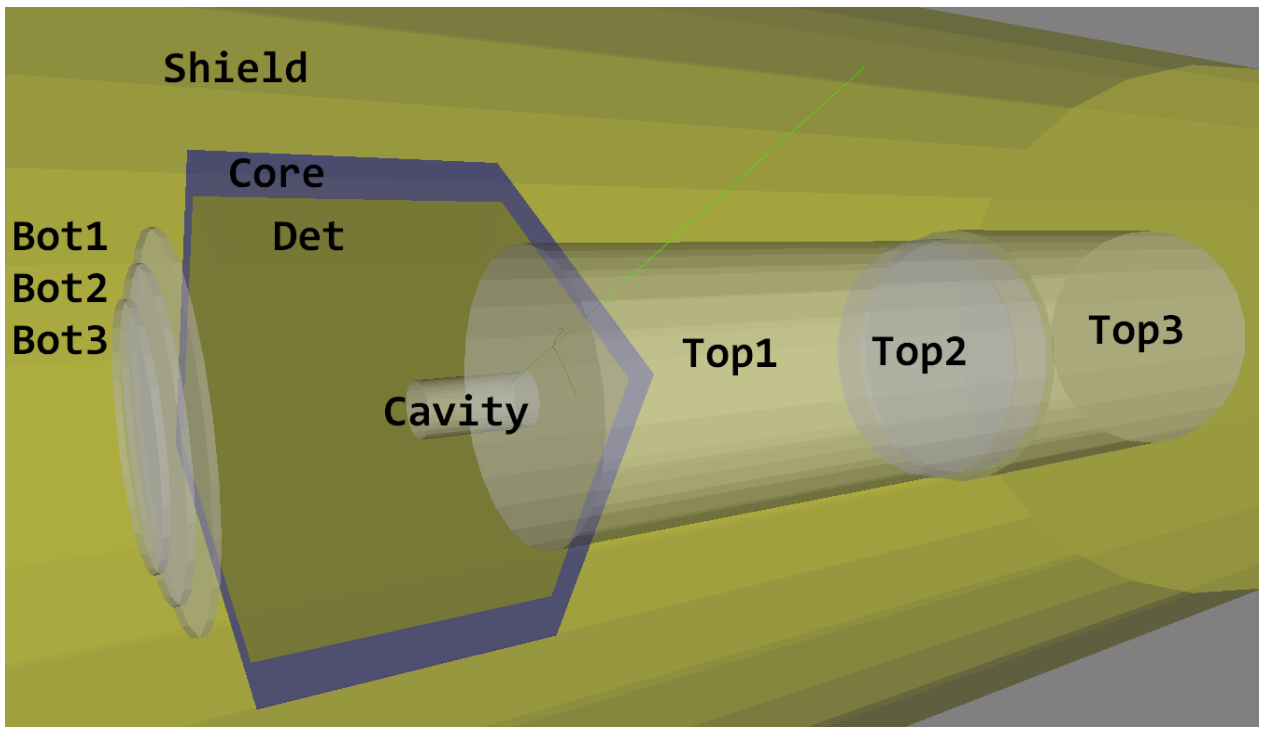

Fig. 3. Example geometry created for simulation with Simourg, including all allowed volumes. A gamma-ray track interacting in the detector is also shown.

The Simourg program is console-based, meaning it does not use a graphical user interface (GUI) and is launched from the command line. For example, on Windows it runs inside a DOS window. The only command-line argument is the name of the configuration file. A sample command to start Simourg from the DOS prompt is:

C:\Simourg>**Simourg_1.exe Simourg_CdWO4.mac >log.txt**

When this command is executed, the program runs with the configuration file `Simourg_CdWO4.mac`, and its standard output is redirected to the file `log.txt`. The verbosity of diagnostic messages sent to standard output is controlled by the `VerboseAll` parameter (range 0–2, default 0)..

To check the geometry, the program can generate interactive 3D vector graphics files in VRML format (see Fig. 3) viewable with any VRML browser or plugin (Cortona, Flux, etc.). To enable this, a parameter `DoPicture 1` should be set in the configuration file (default 0). In this mode, up to 100 VRML files (one per primary gamma) containing volumes and particle trajectories are created. This mode is intended only for geometry debugging; for production runs `DoPicture` must be 0.

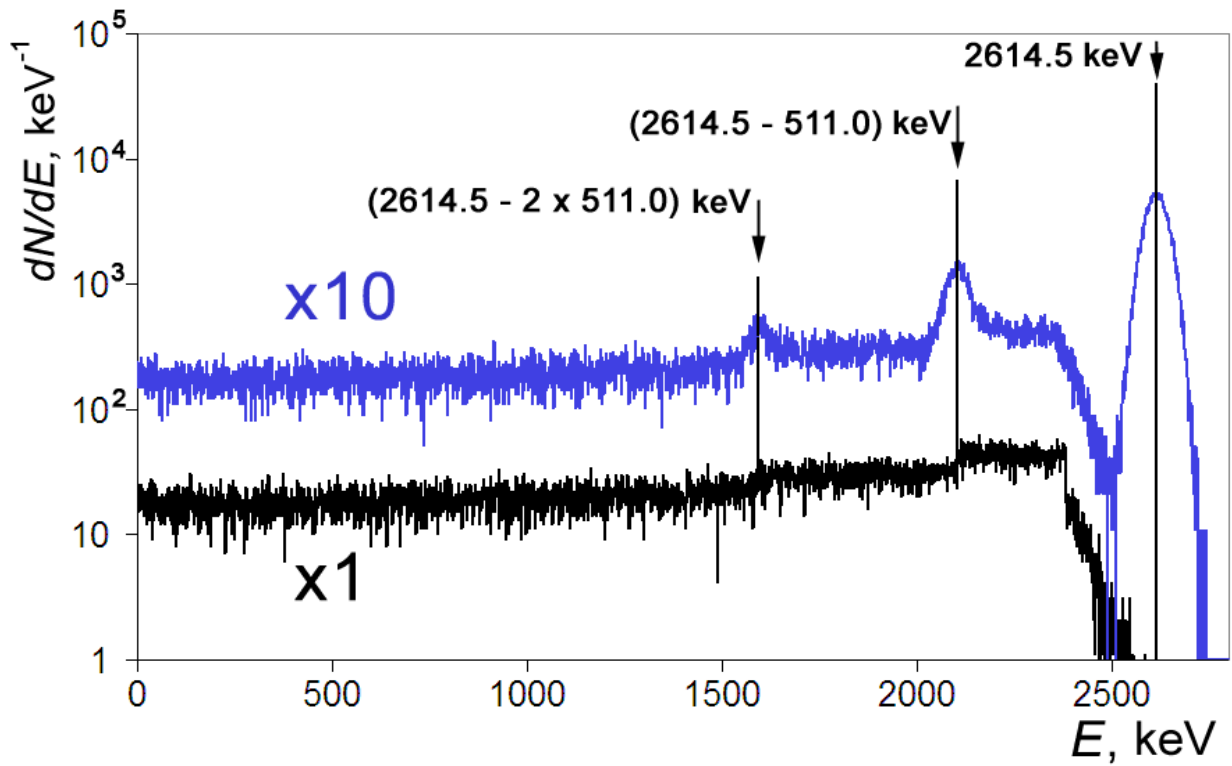

Fig. 4. Spectra obtained for a CdWO$_4$ cylindrical scintillator (Ø50 × 40 mm) and a point source of 2614.5 keV ($^{208}$Tl) gamma rays located 15 mm above the center of the end face. Black line – raw deposited-energy spectrum; blue line – spectrum with finite energy resolution applied (shifted upward ×10 for clarity). Besides the full-energy peak, single-escape and double-escape peaks are visible on the Compton continuum.

Conclusion.

The program Simourg (*Simulator of Usually Requested Geometries*), based on Geant4 and intended for Monte Carlo simulation of nuclear-physics spectrometric gamma detectors with simple geometry typical for studies of long-lived nuclear decays and environmental radioactivity measurements, allows one to obtain detector response functions to monoenergetic gamma rays in the energy range from several keV to several MeV. The distinctive feature of the program is simple control of the simulated setup geometry. The article briefly describes the first version of the program. Future versions are planned to include the possibility of using radioactive decay chains (with electrons, positrons, gamma rays, and alpha particles) generated by the Decay4 program [4] or the Geant4 `RadioactiveDecay` module, thus enabling response functions to real radionuclides (important, e.g., for radionuclides incorporated inside the detector). It is also planned to extend the range of supported geometries by adding prisms with regular polygonal cross-section, cylinders with axial through holes for collimation, and Marinelli beaker shapes.

The work was partially supported by the "Cosmomicrophysics" program of the NAS of Ukraine and the Brain Pool program of the Korean Federation of Science and Technology Societies.